\documentstyle[aps,prl,twocolumn,psfig]{revtex}
\begin{document}

\twocolumn[\hsize\textwidth\columnwidth\hsize\csname@twocolumnfalse\endcsname

\title{\bf Resistivity extrema in double exchange ferromagnetic
 nondegenerate semiconductors }
\author{E.L. Nagaev, A.I. Podel'shchikov, and V.E. Zil'bervarg}
\address{Institute of Radio Engineering and Electronics,
 Russian Academy of Sciences\\
 Mokhovaya ul. 11, korp. 7, Moscow 101999, Russia}

\maketitle

\begin{abstract}

A version of the magnetoimpurity theory of the colossal magnetoresistance
 materials suitable for the double exchange ferromagnetic nondegenerate
 semiconductors is presented. It provides an explanation of the
 nonmonotonic temperature dependence for the charge carrier density
 in them when it displays first a maximum and then a minimum, on
 increase in temperature. Respectively, the resistivity displays first
 a minimum and then a maximum. The theory is based on the relation between
 the charge carrier activation energy and the change in the magnon free
 energy caused by the ionization of an impurity. This is tantamount
 to the relation between the charge carrier density and the so called
 giant red shift of the optical absorption edge.
\end{abstract}
\vskip2pc
]

\section{Introduction}

 The colossal magnetoresistance (CMR) arises in ferromagnetic materials as
 a result of the suppression of the resistivity peak
 by the magnetic field  in the vicinity of the Curie point $T_C$.
 Usually this phenomenon is related to the manganites, though not 
 only they but also all conventional degenerate ferromagnetic
 semiconductors display a similar resistivity peak and CMR
 (see \cite{1}). Still more unusual is the behavior of the
 nondegenerate ferromagnetic semiconductors, in which a 
 resistivity minimum precedes the resistivity peak in the 
 vicinity of $T_C$ (Fig. 1). It is natural to assume that the
 origin of this peak is similar for all ferromagnetic semiconductors
 and manganites. One believes that the manganites are the double 
 exchange systems, i.e., the exchange energy between the localized
 spins and charge carriers far exceeds the bandwidth in them.
 Hence the physical nature of the nonmonotonic temperature dependence
 of the conductivity in doped manganites should be 
 especially close to that in the double exchange ferromagnetic
 semiconductors.

 The resistivity peak in degenerate ferromagnetic semiconductors
 can be explained within the framework of the magnetoimpurity
 theory \cite{1,2,3,4,5}. It is based on the fact that, due to
 the screening, the charge carrier density in the vicinity
 of the ionized donors (acceptors) is higher than far from them.
 The charge carriers give rise to the indirect exchange, which tends
 to support the ferromagnetic ordering. For this reason, at finite
 temperatures the local magnetization in the vicinity of
 impurities is higher than far from them. But the charge
 carrier energy is the lower the higher is the magnetization. For
 this reason, in addition to the Coulomb force, at $T \ne 0$ a
 "magnetic" force appears, which attracts the electron to the
 donor. If the Coulomb force alone is insufficient for the
 electron localization, the total Coulomb + "magnetic" force
 can cause the temperature-induced electron localization of
 electrons, i.e., the transition of the crystal from the
 highly-conductive state to the insulating one.
 
\begin{figure}
\begin{center}
\psfig{figure=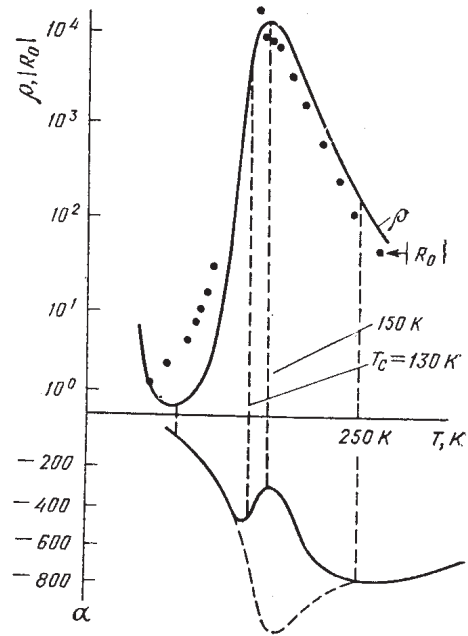,width=87mm}
\caption{Resistivity $\rho$ ($\Omega$ cm), constant of normal
 Hall effect $R_0 = 1/nec$ (cm$^3$/A s), where $n$ is the
 electron density, and the thermoelectric power 
 $\alpha$ (mV/deg) of CdCr$_2$Se$_4$ doped by In (a donor impurity).
 Data on Cd$_{0.99}$In$_{0.01}$Cr$_2$Se$_4$ from \protect \cite{11}
 are presented.
}
\label{Fig.1}
\end{center}
\end{figure}

 On further increase in temperature, the excess local
 magnetization around the donors decreases. Depending on the
 impurity density, two scenarios are possible.\\
 (1) The crystal becomes again highly-conductive. This manifests 
 itself as a high resistivity peak in the vicinity of $T_C$.
 (2) The crystal remains insulating up to the highest temperatures.

 The magnetoimpurity theory is valid for all doped ferromagnetic
 semiconductors, independent of their being or not being double exchange
 and of the existence or nonexistence of the Jahn-Teller effect in them.
 It predicts the colossal magnetoresistance in the cases when the
 impurity density is sufficiently close to that, at which the Mott
 transition takes place \cite{4,5}.

 Some sort of the magnetoimpurity theory should also be valid 
 for the nondegenerate ferromagnetic semiconductors 
 but certainly its specific manifestation in the 
 nondegenerate semiconductors should be different from that 
 in the degenerate ones. Here the magnetoimpurity theory 
 will be formulated for the double exchange semiconductors.

 In addition to the fact that this theory makes it possible
 to explain the properties of some specific ferromagnetic
 semiconductors, it allows one to understand more deeply the
 processes in such double exchange systems as manganites.
 (Let us remind that many of them behave as the degenerate
 semiconductors below $T_C$ and as the nondegenerate ones above $T_C$).
 
 The main results of the present paper can be explained
 qualitatively as follows.
 The drastic difference in the temperature dependence of the 
 resistivity for the ferromagnetic and nonmagnetic
 semiconductors is attributed to the giant red shift of 
 the optical absorption edge specific for the ferromagnetic
 semiconductors. It implies the decrease of the optical gap
 width with decreasing temperature. To explain this 
 phenomenon, it is sufficient to note that the electron
 energy is the lower the higher is the magnetization \cite{1}.

 At very low temperatures, the nondegenerate
 ferromagnetic semiconductor behaves like the nonmagnetic
 one: the charge carrier density increases exponentially with
 temperature. But the activation energy also increases with
 temperature (Fig. 2). The reason for it is
 the fact that the local magnetization in the vicinity of an
 unionized donor is higher than that averaged over the 
 crystal. This is related to the indirect ferromagnetic exchange,
 which is realized by the donor electron. It increases the local
 magnetization close to the donor impurity as compared with its
 mean value over the crystal.  Hence the rate of the
 temperature-induced rise of the donor level is considerably
 less than that of the conduction band bottom. This means that
 the donor level depth first increases with temperature,
 and this process is determined by the band bottom shift
 toward higher energies.

 As the position of the band bottom determines the optical
 gap, one can say that the temperature-induced increase in the donor
 level depth is equal to the temperature increase of the optical gap.
 One calls this shift of the optical gap in the ferromagnetic
 semiconductors "the giant red shift of the optical absorption edge".
 The term "red" is a consequence of the fact that, experimentally,
 this optical shift was observed on decrease in temperature.
 The term "giant" follows from the fact that total optical shift
 amounts to several tens eV, which is by many orders of
 magnitude larger than in nonmagnetic semiconductors.

 The temperature-induced increase in the activation energy first
 slows down the rate of the temperature-induced increase in the
 charge carrier density. On further increase in temperature, a 
 temperature-induced rise in the charge carrier density gives place to
 by a decrease in it. This means that the density passes a 
 maximum and then decreases (Fig. 1). But, on further increase in
 temperature, the long-range ferromagnetic ordering 
 becomes destroyed, and destruction of the local ordering 
 in the vicinity of the donors begins. Then the position of the
 band bottom becomes virtually temperature independent,
 and the donor level depth decreases with 
 increasing temperature. Respectively, after passing a
 minimum, the charge carrier density increases again.

 In reality, this simple physical explanation is valid only 
 qualitatively since the density is expressed not in terms
 of the electron energy averaged over the magnetic state of the crystal
 but in terms of the change in the free energy of the magnetic subsystem
 caused by the impurity ionization. This is the basis for calculations
 carried out below.

 One can point out CdCr$_2$Se$_4$ as an example of the 
 standard ferromagnetic semiconductors with the double 
 exchange. This material can be transferred in the degenerate
 state only by a very heavy doping. This implies a large electron
 effective mass in CdCr$_2$Se$_4$, which is a necessary 
 condition for the double exchange. The double exchange 
 scenario for it corresponds to the appearance of the Cr$^{2+}$ ions
 instead of the regular Cr$^{3+}$. The former play the role of the
 conduction electrons. CdCr$_2$Se$_4$ displays a giant red optical
 shift \cite{6,7,8,9,10} of about 0.15 eV and a nonmonotonic
 temperature dependence of the resistivity of the type described 
 above, i.e. with a minimum and a maximum \cite{11,12}. Figure 1
 in the present paper is taken from \cite{11}.

 \section{Quantum double-exchange Hamiltonian in the spin-wave region}

 Usually in magnetic semiconductors only one type of the
 charge carriers is sensitive to the magnetic ordering (only
 the charge carriers of one sign move over the magnetic cations.
 The charge carriers of the opposite sign move over the nonmagnetic
 ions \cite{1}. For example, in CdCr$_2$Se$_4$ the electrons interact
 strongly and the holes very weakly with the magnetic subsystem.
 For this reason, it is sufficient to take only the former into account.

 The problem of the energy spectrum in the double exchange 
 systems is treated as a single-electron problem. The treatment is based
 on the $s-d$ model with the $s$-electrons modeling the conduction
 electrons or holes moving over the crystal and exhibiting a strong
 exchange interaction with the localized $d$-spins. With accounting for
 the donor (acceptor) impurities the total Hamiltonian of the system is
 written in the form

\begin{equation}
H = H_{sd} + H_{I}.
 \eqnum{1}
 \end{equation}

 The quantum $s-d$ exchange Hamiltonian $H_{sd}$ for double-exchange
 case was deduced by the present author as early as in 1969 (See Refs.
 \cite{1,5,13,14}). Unlike corresponding Hamltonians deduced by other
 authors, this Hamiltonian for the antiferromagnetic $s-d$ exchange 
 is exact to the first order in $AS/W$, and no additional approximations
 (the mean field etc.) were used. To construct it, an 
 effective spin ${\bf S_g}$ of the magnitude $S$ is ascribed to each
 magnetic atom {\bf g}, independently of its being free or occupied
 by a $s$-electron (in the former case it is the true spin). In
 the last case, it formally increases the number of spin degrees of
 freedom by 1 as compared with its actual number. However, the structure
 of the effective Hamiltonian is such that  the contribution from the
 extra degree of freedom vanishes. As for the charge carriers, they are
 treated as spinless fermions with the operators
 $ c ^{*}_{{\bf g}}, c_{{\bf g}}$. This effective Hamiltonian
is:

\begin{eqnarray}
H_{sd}^{ef} = \frac{A(S+1)}{2} \sum c ^{*}_{{\bf g}} 
c_{{\bf g }} - \nonumber\\
t_0 \sum F({\bf S_{g}, S_{g+\Delta}}) c ^{*}_{{\bf g}} 
c_{{\bf g+\Delta }} - 
 \frac{I}{2} \sum {\bf S_{g}S_{g+\Delta}}, \nonumber\\
(2S +1) F({\bf S_{g}, S_{g+\Delta}}) = 
\sqrt{(S + S^{z}_{\bf g}) (S + S^{z}_{\bf g +\Delta})} \nonumber\\
+ \frac{1}{\sqrt{S + S^{z}_{\bf g} }} S^{+}_{\bf g} S^{-
}_{\bf g + \Delta} \frac{1}{\sqrt{ S 
 + S^{z}_{\bf g + \Delta} }} \eqnum{2}
 \end{eqnarray}
 where ${\bf \Delta}$ is the vector connecting the nearest neighbors
 in a simple cubic lattice of the magnetic atoms, $A$ is the
 $s-d$ exchange integral assumed to be negative here (see below),
 $I$ is the $d-d$ exchange integral which is assumed to be positive,
 $t_0$ is the hopping integral, which would have been realized if the
 crystal would have been nonmagnetic. The corresponding band width $W$
 is equal to 12$t_0$. In Eq.(2), small terms of the order of $IS^{2}/t$,
 which  describe a change in the $d-d$ exchange of the atom bearing the
 charge carrier, are disregarded (they are presented in the original papers).
 The first term in Eq.(2) is an additive constant and for this reason
 can be omitted.

 As usually, the inequality $t_0  \gg  IS^2$ must be met as the hopping
 integral $t_0$ is of the first order of magnitude, and  the $d-d$
 exchange integral $I$ of the second order in the small $d$-orbital
 overlapping. In what follows, the condition of the double exchange 
 $W \ll |A|S$ will be also assumed to be met, with the $s-d$ exchange
 integral $A$ being negative. Under these conditions, to a zero
 approximation in $W/AS$, the charge carrier is fixed at one of the
 magnetic atoms, their total spin being $S - 1/2$. This just corresponds
 to the situation in the manganites, where appearance of a hole at a 
 regular ion Mn$^{3+}$ reduces its spin by 1/2.

 It should be pointed out that the case $A < 0$ can be also applied to
 the conduction electrons if the occupancy of the $d$-shell exceeds 5/2.
 Then, due to the Pauli's exclusion principle, the spin of the atom bearing
 a conduction electron should be equal to $S - 1/2$. In this case, the quantity
 $A$ looses the meaning of the Hund's exchange integral and expresses only
 the Pauli's principle. But, as the quantity $A$ enters Eq.(2) only through
 an additive constant in the electron energy, this fact is nonessential. 

 In the case of smaller $d$-shell occupancies, the quantity $A$ for the
 conduction electron is directly related to the Hund's integral and hence
 is positive. One can construct a similar effective Hamiltonian
 also for $A > 0$ by introducing an effective spin $S$ for each atom.
 But in this case one of the states of an atom with the total spin
 $S + 1/2$ remains unaccounted for, and this restricts the applicability
 of this effective Hamiltonian. Nevertheless, it can be used, for example,
 for $2S \gg 1$ or for any spin magnitude in the spin-wave region.
 Then the quantity $F({\bf S_{g}}, {\bf S_{g+\Delta}})$ in Eq.(2)
 should be replaced by its conjugate \cite{1,5,13,14}. As the spin-wave
 region will be considered below, both Hamiltonians for $A > 0$
 and $A < 0$ lead to the same results.

 In the classical limit $S \to \infty$, introducing the polar angles 
 $\theta_{\bf g}, \phi_{\bf g}$ and carrying out a canonical 
 transformation of the spinless operators shifting their phases
 one arrives to the classical Hamiltonian "with the Berry phases"
 deduced for the first time in Refs. [14,1,5].

 The term $H_I$ in Eq.(1) describes the interaction between
 the $s$-electrons and impurities,

\begin{equation}
H_{I} =  \sum _{{\bf j}} V({\bf g - R_j}) 
c^{*}_{{\bf g}}  c_{{\bf g}}, 
 \eqnum{3}
 \end{equation}
 where ${\bf R_{j}}$ are the coordinates of impurity ${\bf j}$.

 Now the quantum double exchange Hamiltonian (2) will be specified
 for the spin-wave region. Introducing the magnon operators
 $b^{*}_{\bf g}, b_{\bf g}$ and performing the
 Holstein--Primakoff transformation in the Hamiltonian (2) 

$$S^{z}_{\bf g} = S - b^{*}_{\bf g} b_{\bf g}, 
 \quad S^{+}_{\bf g} = (2S)^{1/2}b_{\bf g}, \quad S^{-
}_{\bf g} = (2S)^{1/2}b^{*}_{\bf g} $$ 
one obtains the $s-d$ Hamiltonian in the spin-wave region:

\begin{eqnarray}
H_{sd}^{sw} = - t \sum a^*_{\bf g}a_{\bf g+ \Delta} + \nonumber \\
\frac{t}{2S} \sum \left [ \frac{1}{2}(m_{\bf g}+m_{\bf g + \Delta})
 -b^*_{\bf g+ \Delta} b_{\bf g}\right] a^*_{\bf g}a_{\bf g+ \Delta}
 \nonumber \\
+ IS  \sum \left [\frac{1}{2}(m_{\bf g}+m_{\bf g + \Delta}) - b^*_{\bf g+ 
\Delta} b_{\bf g})\right], \eqnum{4}
 \end{eqnarray}
or in the quasimomentum representation

\begin{eqnarray}
H_{sd}^{sw} = - zt \sum \gamma _{\bf k} c^{*}_{\bf k} 
c_{\bf k} \nonumber \\
+ \frac{z t}{4SN} \sum ' (\gamma _{\bf k'} + \gamma_{\bf q - 
q'- k'} - 2\gamma _{\bf q'+k'})c^{*}_{\bf k} c_{\bf k'}
 b^{*}_{\bf q} b_{\bf q'} \nonumber \\
+  \sum \Omega_{\bf q} b^{*}_{\bf q} b_{\bf q}  
\eqnum{4a}
\end{eqnarray}

\begin{equation}
t = \frac{2St_0}{2S+1}, \Omega_{\bf q} = zIS (1 - 
\gamma_{\bf q}), 
\eqnum{5}
\end{equation}

$$\gamma_{\bf k} =  
\frac{1}{z} \sum_{\bf \Delta}{\rm exp}(i{\bf \Delta k}) \qquad 
(z =  6)$$
 where $z$ is the coordination number, the primed sum 
 denotes conservation of the total quasimomentum.

 As is seen from Eq.(5), for $A <0$ even with the complete 
 ferromagnetic ordering, the $s-d$ interaction reduces the
 effective hopping integral by a factor of $1 +1/2S$. For $A >0$
 this  reduction is absent, and $t = t_0$.

\section{Temperature dependence of the conduction band bottom}

 In order to calculate the charge carrier density, first the
 temperature dependence of the conduction band bottom  in the
 spin-wave region will be investigated. The further consideration
 will be carried out under assumption that $1/2S \ll 1$ . One should
 keep in mind that  in the theory of the magnetism the formal
 expansion in $1/2S$ powers usually gives reasonable results even
 for small spins. Then the total Hamiltonian of a perfect crystal
 can be represented in the form

\begin{equation}
H_{sd}^{sw} = \sum E _{\bf k} n_{\bf k} +
 \sum B_{\bf  k,k+q}n_{\bf k}m_{\bf q} + 
  \sum \Omega _{\bf q} m_{\bf q} 
\eqnum{6}
\end{equation}

\begin{eqnarray}
n_{\bf k} = c^{*}_{\bf k} c_{\bf k}, 
m_{\bf q}= b^{*}_{\bf q} b_{\bf q}, \nonumber \\
E_{\bf k} = - zt  \gamma _{\bf k},
B_{\bf k,k+q} = \frac{zt}{2SN}(\gamma_{\bf k}-
 \gamma _{\bf k+q})
\eqnum{7}
\end{eqnarray}

 Introducing the chemical potential $\mu$, one can write for the 
 average number of the $s$ electrons with the quasimomentum {\bf k}:

\begin{equation}
<n_{\bf k}> =  \frac{\partial F }{\partial E_{\bf k}},
\: F = - T {\rm ln}Z_{de} \eqnum{8}
\end{equation}

\begin{eqnarray}
Z_{de} = \prod_{{\bf k, q}}\sum_{\{n,m\}} {\rm exp}\{- 
\beta[(E_{\bf k} - \mu) n_{\bf k} + \nonumber \\
\Omega_{\bf q}m_{\bf q} + B_{\bf k,k+q}n_{\bf k }m_{\bf q}]\} \nonumber \\
= \prod_{{\bf k, q}} \{ \frac{1}{1 - {\rm exp}(- \beta 
\Omega_{\bf q})} + \nonumber \\
\frac {{\rm exp}[\beta(\mu - E_{\bf k})]}{1 
- {\rm exp}[-\beta(\Omega_{\bf q}+B_{\bf  k,k+q })]} \},
\:  \beta = \frac{1}{T}. \eqnum{9}
\end{eqnarray}

 One obtains from Eqs.(8) and (9):

\begin{eqnarray}
<n_{\bf k}> = \nonumber \\
 \left\{1 + {\rm exp}[\beta(E_{\bf k} - \mu)] 
 \frac{\prod_{\bf q}\{1 - {\rm exp}[-\beta(\Omega_{\bf 
 q}+B_{\bf k,k+q })]\}}{\prod_{\bf q} [1 - {\rm exp}(- \beta 
 \Omega_{\bf q})]}\right\}^{-1}
 \eqnum{10}
\end{eqnarray}

 It should be noted that
$$\prod_{\bf q} [1 - {\rm exp}(- \beta \Omega_{\bf q})] = {\rm 
exp}(\beta F^0_m),$$
where 

\begin{equation}
F^0_m = T \sum_{\bf q}{\rm ln}[1 - {\rm exp}(- \beta 
\Omega_{\bf q})] \equiv Nf(J,T) \eqnum{11}
\end{equation}
 is the magnon free energy in the absence of the conduction 
 electrons.

Similarly, introducing the magnon free energy $F_m^{\bf k}$ in 
the presence of the $s$-electron with the quasimomentum ${\bf 
k}$,
\begin{equation}
F_m^{\bf k} = T \sum_{\bf q} {\rm  ln} \{1 - {\rm exp}[- 
\beta (\Omega_{\bf q} +B_{\bf k,k+q})] \} \eqnum{12}
\end{equation}
one can rewrite Eq.(10) :

\begin{equation}
<n_{\bf  k}> = \left \{1 + {\rm exp}[\beta(E^r_{\bf k} - 
\mu)] \right\}^{-1}, \eqnum{13}
\end{equation}
\begin{equation}
E^r_{\bf k}= E_{\bf k} + F^m_{\bf k} - F^m_0,
\eqnum{14}
\end{equation}
 or, using Eq.(11) and keeping in mind
 the fact that $B \sim 1/N$:
\begin{eqnarray}
E^r_{\bf k}= E_{\bf k}+ \frac{zt}{2SN} \sum_{\bf 
q}\gamma_{\bf k}(1 - \gamma_{\bf q}) <m_{\bf q}>, \nonumber \\
<m_{\bf q}> = \frac{1}{{\rm exp}(\beta \Omega_{\bf q})-1}.
\eqnum{15}
\end{eqnarray}

This corresponds to the temperature renormalization of the 
hopping integral:
\begin{equation}
t_{ef}(T) = t\left[1 - \frac{1}{2SN}
 \sum (1 - \gamma _{\bf q}) <m_{\bf q}> \right],
 \eqnum{16}
\end{equation}

 Obviously, the quantity $E^r_{\bf k}$ represents the $s$-electron
 energy renormalized due to the electron-magnon interaction.
 As is seen from Eq.(15), the electron energy increases with
 temperature. As the energy of the holes is assumed to be temperature
 independent, this means that gap $G(T)$ becomes wider.
 On the contrary, the temperature decrease causes its narrowing.
 Using the standard terminology, the temperature shift of the conduction
 band bottom represents the red shift of the optical 
 absorption edge $\delta G(T)$. 
 This shift is proportional to $T^{5/2}$ for $T < J = ISz$
 and to $T$ for $T > J$ (for the applicability of the spin-wave 
 approximation in the latter case, spins should be large,
 $2S \gg 1$). With $z$ = 6, we have 

\begin{eqnarray}
\delta G(T) = G(T) - G(0) = F^m_{\bf k} - F^m_0 \nonumber \\
 = \frac{t \zeta(3/2)}{16 \pi^{3/2}}\left(\frac{6T}{J}\right)^{5/2}, 
\quad {\rm for} \quad T < J, \eqnum{17}
\end{eqnarray}

\begin{equation}
\delta G(T) = \frac{zt T}{2SJ} \quad {\rm for} \quad T >J. \eqnum{18}
\end{equation} 
 where $\zeta(x)$ is the Riemann zeta-function.

 As is seen from Eqs.(17), (18),
 the giant magnitude of the shift is a consequence of the fact
 that it is proportional to the $s$-electron bandwidth. 

 Comparing Eqs.(17) and (15), one sees that two different
 interpretations of the temperature dependence  of the optical 
 absorption edge are possible. The first of them is traditional and
 corresponds to the temperature dependence of the electron energy, which
 is  obtained from Eq.(6) by averaging it over the magnons. Thereby
 an intuitive approach adopted in \cite{1} is confirmed  for the band
 electrons here. But the red shift can also be attributed to the 
change in the magnon free energy due to the electron excitation. In 
considering the impurity conductivity, one will see that only the latter
approach is correct as it makes it possible to describe the temperature
dependence of the electron discrete level correctly (see the paragraph
 after Eq.(27)).

\section {Charge carrier density in an impurity semiconductor}

 Now our task is to calculate the charge carrier density 
 in an impurity semiconductor. The calculation begins with the
 magnon energy spectrum  in the presence of an unionized
 donors. The main feature of the impurity samples is the fact
 that the electron of a unionized donor realizes an indirect
 exchange between $d$-spins. As the electron density diminishes
 exponentially with increasing distance from the impurity, the
 intensity of the indirect exchange diminishes in a similar
 manner.

 For $T = 0$ the electronic wave function can be found exactly in 
 the effective mass approximation when the impurity has 
 is the Coulomb potential. At higher temperatures the 
 increased exchange will increase the magnetization in the 
 vicinity of the impurity. The electron energy is the lower the 
 higher  the local magnetization. Hence an additional force appears 
 which attracts the electron to the donor. Respectively, the effective 
 Bohr radius should decrease, on increase in temperature.
 The problem can be solved by using a variational procedure for the 
 free energy under condition that the donor electron be in the 
 ground state with the hydrogen-like wave function

\begin{equation} 
\psi (r) = \left(\frac{x^3}{\pi a^{3}_{B}}\right)^{1/2} {\rm 
exp} 
\left(- \frac{xr}{a_B}\right), \quad a_B = \frac{2 a^2 t 
\epsilon}{e^2}, \eqnum{19}
\end{equation} 
 where $x$ is the variational parameter and $\epsilon$ is the dielectric 
 constant. The effective Bohr radius $a_B/x$ is assumed to be large 
 comparing with the lattice constant $a$.

 As is seen from the structure of the electron-magnon 
 Hamiltonian (4) and from the form of the wave function (19), an 
 exact treatment of the problem is impossible in this case. To 
 obtain semiqualitative results, it is convenient to replace the 
 nonuniform electron density distribution (19) by an uniform that 
 with an average density $3/4 \pi \rho^3$ inside of the sphere of 
 the radius $\rho$:

\begin{equation}
\rho = \sum g \psi^{2} (g) = \frac{3a_B}{2x},
\eqnum{20}
\end{equation}

 Let us separate such a region from the totality of the magnetic
 atoms, and choose $V$ = const in the Hamiltonian $H_I$ (3)
 in such a way as to ensure the minimal $s$-electron energy
 $E_{I}$ to be equal to that value, which is obtained from the
 Hamiltonian $H_I$ with $V(g) = - e^2/\epsilon g$ with the 
 use of the  trial wave function (19). At $T \to 0$,
 when $x = 1$,

\begin{equation} 
E_{I} \equiv - E_{B} = - e^{2}/2\epsilon a_{B}
 \eqnum{21}
 \end{equation}

 The relative number of the donors $\nu$ is assumed to
 be small. We can divide all regular magnetic atoms into
 those which enter spheres of radius $\rho$ surrounding donors
 and those which are outside these spheres (the number of the
 latter greatly exceeds the total number of the former).

 Now the canonical transformation of the electron operators
 corresponding to sites inside the impurity region will
 be carried out:

\begin{eqnarray}
a_{\bf g} = \frac{1}{\sqrt{N_I}}\sum {\rm exp}(i {\bf pg}) 
a_{\bf p}, \nonumber \\
 b_{\bf g} = \frac{1}{\sqrt{N_I}}\sum {\rm 
exp}(i {\bf qg}) b_{\bf q}, \eqnum{22}
\end{eqnarray}

$$ N_I = \frac{4 \pi \rho^3}{3 a^3} =
 \frac{9 \pi}{2}\left(\frac{a_B}{xa}\right)^3. $$ 

 We will use the expression for the conduction-electron-magnon
 Hamiltonian (6) and retain only terms corresponding to the lowest
 discrete levels in the Hamiltonian $H_{sd} +H_I$ (1), (3)
 with the donor potential $V$ = const. Then we arrive to the 
 following Hamiltonian:

\begin{eqnarray}
 H = (E_I - \mu) \sum n_{Ii}+ \sum (E_{\bf k}- \mu) n_{\bf k} + \nonumber \\
\sum B_{I{\bf p}} n_{Ii}m_{{\bf p}i} +
 \sum B_{\bf k,k+q}n_{\bf k}m_{\bf q} \nonumber \\+
 \sum \Omega_{\bf q}m_{\bf q} + \sum \Omega_{\bf p} m_{{\bf p}i},
 \eqnum{23}
\end{eqnarray}

\begin{equation}
B_{I{\bf p}} = \frac{zt}{2SN_I} (1 - \gamma_{\bf p}),
 \eqnum{24}
\end{equation}
 where $m_{{\bf q},i}$ and $m_{\bf q}$ are the magnon operators for
 the $i$-th sphere and outside the spheres, which surround donors,
 respectively. Since the magnon number operators for different 
 donor regions and outside them are constructed of magnon operators
 $b^{*}_{\bf g}$ and $b_{\bf g}$ with different ${\bf g}$, all
 the operators $m_{{\bf q},i}$ and $m_{\bf q}$ are independent.

 Further, $n_{I,i}$ and $n_{\bf k}$ are the operators for an 
 electron in the localized state at the donor $i$ and for the delocalized
 electrons outside the spheres with the quasimomentum ${\bf k}$,
 respectively.

 It should be noted that Eq.(23) is also valid for degenerate semiconductors
 and leads to the well-known equation for the magnon frequencies in them

$$\omega_{\bf q} = \Omega_{\bf q} +
 \frac{zt}{2SN}\sum_{\bf k} \gamma_{\bf k}
(1 - \gamma_{\bf q}). $$
 One ascribes usually this equation to Furukawa \cite{15}, though in
 reality it was first obtained by the present author \cite{13,1}.

 The mean number of electrons at a donor is calculated in the
 same manner as in Eqs.(8) to (14). It is given by the expression
 (the index of the donor is omitted):

\begin{equation}
<n_{I}> = \{1 +{\rm exp}[(E_{I} +\delta F_{mI}(T) -\mu)/T] 
\}^{-1},
\eqnum{25}
\end{equation}

\begin{equation}
 \delta F_{mI}(T) = F_{mI} - F_{mI}^{0} 
 = N_{I} [f(J_I,T) -  f(J,T)]
 \eqnum{26}
\end{equation}
 where $F_{mI}$ and $F^{0}_{mI}$ are the magnon free energies
 for a region of radius $\rho$ containing an unionized and ionized
 donor, respectively. The magnon free energy $f(J,T)$ is given
 by Eq.(11) and the quantity $f(J_I,T)$ differs from it by the
 replacement of the $d-d$ superexchange integral $J$ by 
 the total impurity exchange integral $J_I$. The latter includes the 
 contribution from both $d-d$ superexchange and indirect exchange via
 the $s$-electron (24):

\begin{equation}
J_I = J + \frac{zta^3 x^3}{9 \pi S a_B^3}.
\eqnum{27}
\end{equation}

 Obviously, the quantity $\delta F_{mI}$ describes the temperature
 shift of the donor level. This shift differs strongly from that
 obtained from a Hamiltonian of the electron-magnon interaction
 similar to Eq.(6) by its averaging over the magnons. Really, as
 the quantity $B_{I{\bf p}}$ (24) is not asymptotically
 small, one cannot restrict oneself to a linear approximation
 in it. Hence, a generally correct interpretation of the
 temperature shift of the donor level is as follows: it is the 
 difference between the magnon free energies resulting from 
 the donor  ionization. This resolves the dilemma formulated
 in the end of Section 3.

 Equating the number of ionized donors determined from Eq.(25)
 to the total number of the conduction electrons (13), we find an
 expression for the charge carrier density $n_{cc}$ for the quadratic
 dispersion relation $E_{\bf k} = t k^2a^2 = k^{2}/2m$:
\begin{equation}
n_{cc} = (n n_{eff})^{1/2}{\rm exp}[E_A(T)/2T], 
n_{eff} = \frac{(mT)^{3/2}}{2 \sqrt{2 \pi}} 
 \eqnum{28}
\end{equation}
$$E_A(T) = E_{I}+ \delta(T), \qquad \delta(T) =  \delta F_{mI}(T) -  
\delta G(T),$$
 where $n_{eff}$ is the effective density of states in the conduction
 band, and $n =  \nu/a^{3}$ is the donor  density. It should be 
 recalled that $E_I(0) = - E_B$ is negative.

It can be ascertained that the activation energy for the charge carrier
 density $[-E_A(T)/2]$ in Eq.(28) increases with temperature in
 the spin-wave region (Fig. 2). First, it should be noted that for
 very large $a_B$ the temperature dependence of $E_A$ disappears since
 the expansion of $f(J_I,T)$ in the powers of $1/a_B^3$ gives an
 expression for $\delta F_{mI}$, which coincides with $\delta G(T)$.
 For smaller $a_B$, if  the second term in Eq.(26) dominates,
 then one can neglect  $F_{mI}$ in Eq.(26),  i.e.
 $\delta (T) = -N_I f(J)  - \delta G(T)$.

\begin{figure}
\begin{center}
\psfig{figure=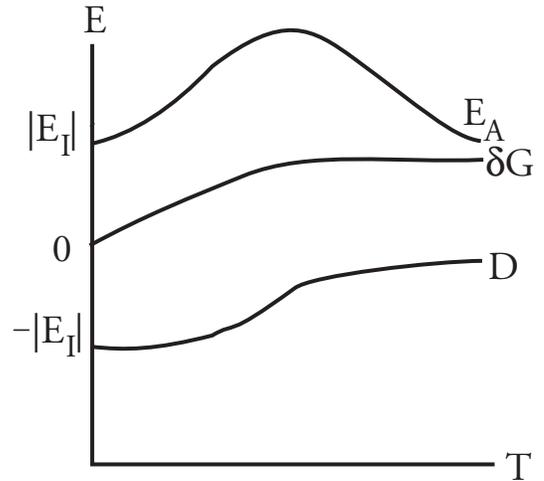,width=87mm}
\caption{Qualitative temperature dependences of the conduction
band bottom $\delta G$ (it corresponds to the 
giant red shift of the optical absorption edge); of donor level 
$D = E_I +\delta F_{mI}$; of the doubled activation energy for the
charge carrier density and conductivity $|E_A|$.}
\label{Fig.2}
\end{center}
\end{figure}

 According to Eq.(11), the first term, which describes the temperature
 shift of the donor level, is positive and has a value of order
 $N_I J (T/J)^y$. The quantity $y(T)$ decreases from 5/2 to 1
 with increasing temperature. According to Eqs.(17), (18), the second
 term, which describes the optical gap shift, is negative and of the
 order of $(W/S)(T/J)^y$.

 As $W \gg JS$, the quantity $\delta (T)$ should be negative with its 
 absolute value increasing with temperature. This means that the 
 activation energy in Eq.(28) increases with temperature. In this 
 case, its temperature dependence is almost completely 
 determined by the shift of the optical absorption gap $\delta G(T)$.

 Physically, this effect can be explained as follows. The donor 
 electron realizes the ferromagnetic indirect exchange in the 
 vicinity of the impurity. For this reason the ferromagnetic 
 coupling there is stronger than on the average over  the 
 crystal. Hence, with increasing temperature, the local  
 ferromagnetic ordering degree in the donor vicinity decreases more 
 slowly than on the average over the crystal. But the electron 
 energy is the lower the higher is the degree of the  ferromagnetic 
 ordering. As a result, the donor level depth increases with the 
 temperature in the spin-wave region. This fact can cause 
 appearance of a maximum in the charge carrier density at a 
certain temperature $T_m$. According to Eq.(28), its condition 
is
\begin{equation}
E_I + \delta (T_m) = T_m \frac{d \delta (T_m)}{d T}.
\eqnum{29}
\end{equation}

 It should be noted that some experimentalists \cite{16,17} pointed out
 that the activation energy in the ferromagnetic semiconductors must be 
 temperature-dependent, and a condition of the type of Eq.(29) was proposed
 for the density maximum. But no expression for the quantity $\delta$ was 
 obtained by them.

 Below the explicit formula for this quantity will be used.
 If, in accordance with Eqs.(17), (18), in the low
 temperature range one puts $\delta \sim WT^y/SJ^y$ with $y = 5/2$,
 then 
 \begin{equation}
T_m  \sim  J \left[\frac{E_BS}{W}\right]^{2/5},
\eqnum{30}
\end{equation}
 where $E_B$ is given by Eq.(21). As $W  \gg  E_B$, this means that the
 maximum charge carrier density is reached still in the spin-wave region
 below $T = J$ in accordance with the choice of $n$ made above. It
 manifests itself as a resistivity minimum.

 This maximum assumes existence the subsequent charge carrier density
 minimum, as at elevated temperatures the ferromagnetic ordering is destroyed,
 and  $<n_{cc}(T)>$ should increase with temperature exponentially (Fig. 1).
 This is nothing else as a qualitative proof of the existence of the
 resistivity peak at temperatures higher than the resistivity 
 minimum. Physically, such a nonmonotonic behavior of $<n_{cc}(T)>$ can
 be explained by the fact that, on increase in temperature, after
 destruction of the ferromagnetic ordering  far from impurities,
 its destruction close to the impurities begins. Hence the donor level rises.
 As the conduction band bottom band remains fixed at such 
 temperatures, the donor level depth decreases, and the rate of the 
 temperature growth for the $<n_{cc}(T)>$ increases (Fig. 2). But, as these 
 processes take place at temperatures of order of the Curie 
 point, their analytical treatment is hardly possible at present.

 In particular, at temperatures comparable with the Curie point, one 
 should take the ferron effect into account: the electron is dragged
 in by the region of the enhanced magnetization and simultaneously
 supports it, realizing the ferromagnetic indirect exchange inside
 it. This process decreases the donor free energy and hence the charge
 carrier density. In the paramagnetic region, such a state of a donor was
 investigated in \cite{18,1} for non-double exchange systems.
 For the double exchange systems it is not investigated yet. But in
 the spin-wave region one can take the ferron effect into account if one considers  temperatures sufficiently low, when the term $\delta F_{mI}$
 in the total free energy can be considered as a perturbation.
As follows from Eqs.(3), (19), (26), the free energy of a system
 containing a donor has the form
\begin{equation}
F_I(x)=E_B(x^2-2x)+\delta F_{mI}.
\eqnum{31}
\end{equation}

Then the optimum value of $x$ is
\begin{equation}
$$x = 1 - \frac{1}{2} \frac{d \delta F_{mI}(1)}{dx}.
\eqnum{32}
\end{equation}

 The free energy coincides with the quantity $F_I(1)$ virtually used
 above up to the second order in ($x-1$). The parameter $x$ increases
 with temperature, and the electron orbital radius $a_B/x$  decreases as
 should be the case.

\section*{Acknowledgements}

 This work was supported in part by Grant No. 01-02-97010
 of the Moscow Region Government - Russian Foundation for Basic Research,
 by Grant INTAS-97-open-30253, and by the agreement with the Russian
 Ministry of Science and Industry.


\end{document}